\documentclass[aps,superscriptaddress,showpacs]{revtex4}
\usepackage{amssymb}
\begin{document}
\font\hn = phvrrn
\title{A stable static Universe?}
\author{Carlos Barcel\'o}
\email[]{carlos@iaa.es}
\affiliation{Instituto de Astrof\'{\i}sica de Andaluc\'{\i}a, 
Camino Bajo de Hu\'etor N24, 18008 Granada, Spain.}
\author{G.E. Volovik}
\email[]{volovik@boojum.hut.fi}
\affiliation{Low Temperature Laboratory, Helsinki University of Technology,
Box 2200, FIN-02015 HUT, Finland and L.D. Landau Institute for Theoretical
physics, 117334 Moscow, Russia.} 

\date{Version 3; 14 July 2004; \LaTeX-ed \today}
\begin{abstract}
  
Starting from the assumption that general relativity might be an
emergent phenomenon showing up at low-energies from  
an underlying microscopic structure, we re-analyze the
stability of a static closed Universe filled with radiation. In
this scenario, it is sensible to consider the effective
general-relativistic configuration as in a thermal contact
with an "environment" (the role of environment can be played, for example, by
the higher-dimensional bulk or by the trans-Planckian degrees of
freedom). We calculate the free energy at a fixed temperature of this
radiation-filled static configuration. Then, by looking at the free
energy we show that the static Einstein configuration is stable
under the stated condition.

\end{abstract}
\widetext
\maketitle
\def\g{\kappa}
\def\half{{1\over2}}
\def\L{{\mathcal L}}
\def\S{{\mathcal S}}
\def\d{{\mathrm{d}}}
\def\x{{\mathbf x}}
\def\v{{\mathbf v}}
\def\im{{\rm i}}
\def\etal{{\emph{et al\/}}}
\def\det{{\mathrm{det}}}
\def\tr{{\mathrm{tr}}}
\def\ie{{\emph{i.e.}}}
\def\bnabla{\mbox{\boldmath$\nabla$}}
\def\Box{\kern0.5pt{\lower0.1pt\vbox{\hrule height.5pt width 6.8pt
    \hbox{\vrule width.5pt height6pt \kern6pt \vrule width.3pt}
    \hrule height.3pt width 6.8pt} }\kern1.5pt}
\def\HRULE{{\bigskip\hrule\bigskip}}
\def\be{\begin{equation}}
\def\ee{\end{equation}}
\def\implies{\Rightarrow}

\section{Introduction}

The development of a geometrical description of the gravitational
field led Einstein in 1917 to propose that the Universe could be, in
the overall, a three-dimensional sphere with no other evolution than
that provided by local physics \cite{einstein}. To make this sort of
equilibrium state for the Universe compatible with his geometrical
field equations, he introduced the afterward famous cosmological
constant.  The cosmological constant succeeds in counterbalancing the
collapsing tendency of all the matter in the Universe.  Later, in
1930, Eddington proved that Einstein's static Universe was unstable
under homogeneous departures from the equilibrium state
\cite{eddington}.
At that time, the recession of galaxies had already been observed by Hubble.
For this reason and in view of the instability, the Einstein model was
considered as a possible initial state for the Universe that once
destabilized would start to expand. (This point of view has been 
revisited recently in \cite{ellis,barrow}).

Eddington did not clearly analyze what could trigger the development
of the instability, but vaguely associate it with the formation of
condensations.  Thanks to a very interesting series of works
\cite{lemaitre,bonnor,harrison,gibbons,ellis,barrow}(we do not intend to be 
exhaustive), at present we know that the instability of 
 models of the Universe that are closed, and homogeneous, 
isotropic and static on the overall, is a much more subtle issue. 
 Here on we will call all these models Einstein models, independently 
of their matter content, although the model propossed by Einstein 
himself correspond to a Universe filled with dust.

On the one hand, Harrison showed
\cite{harrison} that if the equation of state of matter is such that
its associated speed of sound $c_s$ is greater than $1/\sqrt{5}$ all
the physical inhomogeneous perturbations are neutrally stable. This
has been recently emphasized and extended by Ellis and Maartens
\cite{ellis} and Barrow \etal \cite{barrow}.
This case includes a radiation dominated universe.  For $0 \neq c_s <
1/\sqrt{5}$ only a finite set of inhomogeneous modes become
unstable. What happen is that in a finite-size Universe, as  an
Einstein model, the Jeans scale for the formation of condensations is
a significant fraction of the maximum attainable scale. Therefore for
higher enough speeds of sound, only the Universe as a whole could
develop an instability. For this same condition, $c_s > 1/5$, Gibbons
also showed that the Einstein point correspond to a local maximum of
the entropy among the set of geometries conformally related to it that
have a moment of time symmetry \cite{gibbons}.

On the other hand, Bonnor showed \cite{bonnor} that, at least in the
simplest case in which matter satisfies an equation of state, in order
to really depart from the Einstein state one would need a global
decrease in pressure in the entire Universe, 
(this process had already been discussed by
Lema\^itre in 1931 under the name of ``stagnation''
\cite{lemaitre}). This suggests that a static Universe describable 
in the cosmological scales as filled with dust, for which $p=0$, could
not be able to change its static global state, but only develop
instabilities on smaller scales. On the other extreme, a static
Universe filled with radiation could in principle exit from this state
towards a Friedman expansion by decreasing its pressure. Here we will
concentrate on this later model and its instability.

In this letter, we will analyze
the stability of a closed and static Universe filled with
radiation,  but starting from notions somewhat different from
those in standard general relativity.
General relativity is commonly considered to be a low-energy effective
theory that emerges from a deeper underlying structure. A particular
realization of this situation is suggested by the gravitational
features showing up in many condensed matter system (such as liquid
Helium) in the low-energy corner \cite{volovik-book}. These type of
systems suggest that both, matter particles and interaction fields,
could be different emergent features of the underlying system: they
will correspond to quasiparticles and collective-field excitations of
a multi-particle quantum system.  For example, in the phase A of
$^3$He, the quasiparticles correspond to Weyl fermions and the
collective fields to electromagnetic and gravitational (geometrical)
fields.

Imagine now that a Universe of the Einstein type was the effective
result of describing the geometric and matter-like degrees of freedom
emerging from the underlying structure.  A photons-filled Einstein
Universe will have a specific temperature.  In the standard general
relativity, the stability of the system is analyzed under the
assumption of adiabaticity: there is no heat transfer in or out the
Universe because ``there is nothing outside the Universe''.  However, in
the emergent picture described above there is not any a priory reason
to consider the system as effectively thermodynamically closed (let us
remark that this is a non standard general relativitic
behaviour). Therefore, it is natural to ask what would happen when
perturbing the Einstein state if the temperature of the underlying
structure stays constant.  We will not enter on what sets and controls
this temperature, but only assume that it is independent from the
behaviour of the effective Universe. Let us emphasize that we do
not disturb the general relativity, so that the solution for the
equilibrium static Universe is the same that follows from the Einstein
equations. We only allow for the heat exchange with the
``environment''.

Apart from the emergent gravity picture, there exist other situations
in which the image of an externally fixed temperature might also have
sense. These are situations in which the 4-dimensional world of
standard general relativity does not conform a completely closed
system.  We can imagine, for example, scenarios with extra dimensions,
(playing the role of environment), from which energy can flow in and
out the 4-dimensional section.

In the following we will compare Eddington's stability analysis with
an analysis based on the fact that the temperature of the system is
kept fixed. For that, we will calculate the free energy of radiation
dominated Einstein states. Let us start now by reviewing the standard
Eddington instability argument.

\section{Eddington's instability  analysis}

Consider a generic positive-curvature FRW metric written in the form 
\begin{eqnarray}
ds^2=-N^2(t)dt^2+a^2(t)\Omega_{ij}dx^idx^j. 
\end{eqnarray}
Here $N(t)$ is the lapse function, $a(t)$ the scale factor and
$\Omega_{ij}$ the metric on a unit three sphere. Following Schutz
\cite{schutz}, the Einstein equations for a Universe filled with a
perfect fluid, can be obtained by varying the action
\begin{eqnarray}
S={1 \over 16 \pi G}\int d^4x \sqrt{-g} \; (R-2\Lambda)
+\int d^3x \sqrt{-g} \; K
+\int d^4x \sqrt{-g} \; p.
\end{eqnarray}
This is the standard Einstein-Hilbert action supplemented with a
boundary term and the volume integral of the fluid pressure $p$. To
obtain the standard form of the Friedman and Raychaudhuri equations we
can substitute the previous FRW ansatz in the action and, after
variation, set the lapse function equal to unity, $N=1$.
Specifically the action can be written as
\begin{eqnarray}
S={2\pi^2 \over 8 \pi G}  \int dt \; Na^3 \; 
\left[3\left(-{\dot a^2 \over a^2} {1 \over N^2}
+{1 \over a^2}\right) -\Lambda \right] 
+2\pi^2 \int dt \; Na^3 \; p(N).
\label{action}
\end{eqnarray}
Here, the explicit dependence of the pressure on its argument
is determined by the condition
\begin{eqnarray}
N{\partial p(N) \over \partial N }=-(\rho+p).
\end{eqnarray}
For example, for a radiation equation of state $\rho=3p$ one obtains
from this condition that $p=C N^{-4}$ with $C$ a constant.

By looking at the previous action and having in mind that we are interested 
on the analysis of the static solutions of the system, we can define 
a different and simpler functional containing all the relevant information:
\begin{eqnarray}
S_{\rm st}=\int dt \; Na^3 \; 
\left[
{3 \over a^2} -\Lambda 
+ \tilde p(N)
\right]. 
\label{S-function}
\end{eqnarray}
Here, we have rescaled the density and pressure as $\tilde \rho= 8\pi
G \rho$, $\tilde p= 8\pi G p$.
We can easily see that by varying with respect to $N$ and $a$ and 
setting $N=1$ we obtain
\begin{eqnarray}
&&(\Lambda +\tilde \rho)a^2=3 \\
&&(\Lambda -\tilde p)a^2=1.
\end{eqnarray}
In the case of a Universe filled with radiation, $\tilde\rho=3\tilde p$, 
these relations
give us the radiation Einstein conditions 
$\tilde \rho=\Lambda$, $a^2_0=(3/2) \Lambda^{-1}$. 


By looking at the functional $S_{\rm st}$,(setting $N=1$), one can
also see that the Einstein point is not stable. Taking into account
that the kinetic term for the scale factor enter the gravitational
action (\ref{action}) with a negative sign, the local maxima of the 
functional $S_{\rm st}$ will correspond to unstable points.
This is just the case for the Einstein point (one can perform
explicitly the second variation with respect to $a$ to check this
local behaviour). This is the Eddington instability we described in
the introduction. We have obtained it here in this variational way 
because of later convenience in comparing it with the result
obtained from the free energy function.

\section{Free energy of a static gravitational configuration}

Let us now take a completely different point of view. Let us analyze
what happen when the perturbation to the Einstein model is performed
as immersed in a thermal reservoir at a fixed temperature.  For that
let us consider the free energy of static gravitational
configurations. We follow the treatment described in \cite{zelnikov}
and references there in.  The free energy of the purely gravitational
part of a static configuration is determined by the Euclidean action
of the configuration (see for example \cite{york}):
\begin{eqnarray}
F_0=T_0I=-{T_0 \over 16 \pi G}\int d\tau d^3x \sqrt{g_e} \; (R_e-2\Lambda).
\end{eqnarray}
Here the temporal coordinate $\tau$ is periodic with period equal to
the inverse of the temperature $T_0$. The symbols $R_e$ and $g_e$
stand respectively for the Euclidean curvature and Euclidean metric of
the configuration. We can realize that this term coming from the
purely gravitational sector do not actually depend on the temperature
so they will be there even at zero temperature
\begin{eqnarray}
F_0=-{1 \over 16 \pi G}\int d^3x \sqrt{g_e} \; (R_e-2\Lambda).
\end{eqnarray}
We are assuming that a proper Einstein-Hilbert behaviour is emerging
in the low-energy corner. Let us remain you that this is not what
normally happen in the standard condensed matter systems we know of. In
these cases the Einstein-Hilbert behaviour is supplemented with
non-covariant terms (see \cite{volovik-book}).

Let us now consider the free energy of a gas of photons
(radiation) inside a curved but static geometry. The leading term in
the temperature on the free energy function is
\begin{eqnarray}
F_1=-{\sigma \over 3}\int d^3x \sqrt{g_e} T^4(x),
\end{eqnarray}
where $\sigma \equiv \pi^2 k_B^4 /  15 \hbar^3 c^2 $ is the 
Stefan-Boltzmann constant, and
\begin{eqnarray}
T(x)={T_0 \over \sqrt{g_{00}(x)} }
\end{eqnarray}
the Tolman temperature; (we will see later that there are other
contributions to the free energy in lower powers of the
temperature). For the particular geometries we are interested in
here, the total free energy can be written as 
\begin{eqnarray}
F(a,N,T)=F_0+F_1=
{2\pi^2 a^3 \over 8 \pi G}
\left[-{3 N \over a^2} + N \Lambda  -{\tilde p \over N^3} \right],
\label{F-radiation}
\end{eqnarray}
with 
\begin{eqnarray}
\tilde p={1 \over 3}\tilde \rho=
{8 \pi G \over 3}\sigma T_0^4: {\rm constant}. 
\end{eqnarray}

>From this free energy, associated with a static geometry filled with
radiation at a temperature $T_0$, we can obtain the Einstein 
static condition. It corresponds to the one that extremises the function $F$.
Variation with respect to $N$ with an afterwards evaluation in $N=1$
gives 
\begin{eqnarray}
(\Lambda + 3 \tilde p)a^2 = (\Lambda + \tilde \rho)a^2 = 3.
\end{eqnarray}
Variation with respect to $a$ yields
\begin{eqnarray}
(\Lambda - \tilde p)a^2 = 1.
\end{eqnarray}
Therefore we have found the same expressions that before: the 
conditions for a static Einstein Universe filled with radiation.

By inspection of the free energy function we can see that now the
Einstein static point is located at a local minimum.  This is the main
point we want to highlight in this letter.  {\em If the perturbation
of the  radiation filled Einstein  Universe were done under
the influence of an externally fixed temperature, (something
outside the realm of standard general relativity) then, the Einstein
point will be stable}.

\section{Corrections due to the temperature}

In the previous section we analyzed the free energy of a system
composed by static geometries of the Einstein type, as the containers,
plus photon gases, as the contents. The free energy
(\ref{F-radiation}) is an approximation as it does not contain
additional contributions in smaller powers of the temperature. In the
high-temperature limit $T^2 \gg \hbar^2 R_e$, the total free energy for
a gas of photons in a static spacetime is \cite{zelnikov,gusev}
\begin{eqnarray}
F=
-{1 \over 16 \pi G}\int d^3x \sqrt{g_e} \; (R_e-2\Lambda)
-{\sigma \over 3} \int d^3x \sqrt{g_e} T^4
+\bar \sigma \int d^3x \sqrt{g_e} T^2 [R_e+6\omega^2]
\end{eqnarray}
with
\begin{eqnarray}
\omega_\mu = {1 \over 2} \partial_\mu \ln |g_{00}(r)|.
\end{eqnarray}
Here the prefactor $\bar \sigma$ in the $T^2 R_e$ term is $\bar \sigma
= {N_v \over 36 \hbar }$ and is obtained by integration over thermal
photon fields (see \cite{altaie}). If the integration had been made
using minimally coupled scalar fields the coefficient would have been
$\bar \sigma =-{N_s \over 144 \hbar }$ with $N_s$ the number of
minimally coupled scalar fields; equivalently for $N_d$ Dirac fermions
one would have $\bar \sigma = {N_d \over 144 \hbar }$
\cite{zelnikov,gusev}.  
In our particular case, this free energy yields
\begin{eqnarray}
F(N,a,T_0)=
{2\pi^2 a^3 \over 8 \pi G}
\left[
-{3 N \over a^2} + N \Lambda  -{\tilde p \over N^3} 
+8 \pi G \bar \sigma {6 T_0^2 \over N a^2}
\right],
\end{eqnarray}
or expressing everything in terms of $\tilde \rho$ and 
denoting the constant factor $({8 \pi G / \sigma})^{1/2} 6 \bar \sigma$
by the letter $b$, 
\begin{eqnarray}
F(N,a,T_0)=
{2\pi^2 a^3 \over 8 \pi G}
\left[
-{3 N \over a^2} + N \Lambda  -{\tilde \rho \over 3N^3} 
+b \;
{ \tilde \rho^{1/2} \over N a^2}
\right].
\end{eqnarray}
Varying with respect to $N$ and $a$ we find now 
\begin{eqnarray}
&&(\Lambda + \tilde \rho)a^2 -
b
\tilde\rho^{1/2} = 3, \\
&&
\left(\Lambda - {1 \over 3} \tilde \rho \right) a^2
+{1 \over 3}b 
\tilde\rho^{1/2} = 1.
\end{eqnarray}
Manipulating these two conditions one obtains%
\begin{equation}
 \Lambda=  {3 \over 2a^2}={\tilde \rho\over
{1 + {2\over 3} b\tilde \rho^{1/2}}}.
\label{Lambda}
\end{equation}
Note that in emergent gravity the external
temperature and thus $\tilde \rho$ is fixed, while the cosmological
constant $\Lambda$ (i.e. the vacuum pressure) is adjusted to the
thermodynamic equilibrium. This is the reason why in the emergent gravity
the cosmological constant is always much smaller than its `natural'
Planck value: in our case $\Lambda\sim T_0^4/E_{\rm Planck}^2\ll E_{\rm
Planck}^2$.

In the particular case of photon gas with $\bar \sigma = {1 \over 36
\hbar }$ one obtains the following modification of the Einstein point
by thermal fluctuations (here we use $\hbar=c=1$):
\begin{equation}
\Lambda=  {3 \over 2a^2}={8\over 15}{\pi^3  GT_0^4\over 1 
+ (8/9)\pi G T_0^2}~.
\label{LambdaT}
\end{equation}
This corresponds to the original Einstein point with
 the Newton constant renormalized by thermal fluctuations:
\begin{equation}
\Lambda=  {3 \over 2a^2}={8\over 15}\pi^3  \tilde G T_0^4~,~\tilde G^{-1}=
G^{-1} + {8\over 9}\pi T_0^2~.
\label{Renormalization}
\end{equation}
For $GT_0^2\ll 1$, i.e. when $T_0^2\ll E_{\rm Planck}^2$, the result
in Eq.(\ref{LambdaT}) coincides with that obtained by Altaie and
Dowker, see Eqs.(44) and (41) in \cite{altaie}: \begin{equation} {1
\over a^2}={16\over 45}\pi^3 GT_0^4 \left(1 -{8\over 9}\pi G
T_0^2\right)~.  \label{Dowker}
\end{equation}
This demonstrates that in equilibrium and in the limit $T_0^2\ll
E_{\rm Planck}^2$, the thermal correction to Einstein point which
follows from minimization of the free energy coincides with the result
following from the conventional general relativity approach, in which
the Einstein equations are solved in a self-consistent manner taking
into account thermal fluctuations \cite{altaie}.

Although our treatment here cannot consider the dynamics of the system, we
can realize that general equations for the evolution of an effective FRW
Universe in contact with a fixed temperature reservoir will be  different
from that of Einstein. This in particular must include the evolution of
the `cosmological constant'
 to its equilibrium value.

\section{Homogeneous but anisotropic perturbations}

In standard general relativity the Einstein point for a Universe
filled with radiation is also unstable against homogeneous but
anisotropic perturbations of the metric of the Bianchi type IX (see
for example \cite{barrow}). For completeness, in this section we want
to see whether the Einstein isotropic point is, on the contrary,
stable in our approach.

In order to analyze the stability of the Einstein state from our
emergent gravity point of view, we need to calculate the free energy 
of static configurations of the Bianchi IX type, which include the Einstein
isotropic configuration. The general metric for these models is \cite{misner}
\begin{eqnarray}
&&ds^2=-N dt^2 + \sum_{n=1}^{3} a_n^2 \sigma_n^2, \\
&&\sigma_1 = \sin \psi d\theta - \cos \psi \sin \theta d \varphi, \\
&&\sigma_2 = \cos \psi d\theta + \sin \psi \sin \theta d \varphi, \\
&&\sigma_3 =-(d \psi + \cos \theta d\varphi ).
\end{eqnarray}
Now, the free energy for these configurations results  
\begin{eqnarray}
F(a_1,a_2,a_3,N,T)=
{2\pi^2 a_1 a_2 a_3 \over 8 \pi G}
\left[
{N a_1^2\over a_2^2 a_3^2} +{N a_2^2\over a_1^2 a_3^2} +
{N a_3^2\over a_1^2 a_2^2} 
-{2N \over a_1^2}-{2N \over a_2^2}-{2N \over a_3^2} 
+ N \Lambda  -{\tilde p \over N^3} \right].
\label{F-bianchi9}
\end{eqnarray}
Again, it is not difficult to see that the Einstein point is an
extremum of this free energy $a^2_1 = a^2_2 = a^2_3 = 
(3/2) \Lambda^{-1} = (3/2) \tilde \rho^{-1}$, and that it is 
a local minimum.

\section{Discussion}

The Einstein static model was introduced as an state of equilibrium
for the Universe as a whole. A few years after its introduction it was
pointed out that it will be unstable under perturbations. At present
we know that the issue of stability of  closed and static
configurations is not as clear cut as it was considered, 
although there are still several ways to instability.
For example, a radiation dominated Universe, could be driven
globally unstable by a sudden decrease in pressure in the entire Universe.

Here, we have analyzed this global instability  of a static
Universe, in the simple case of radiation dominance, but taken a
different perspective on the essence of gravity. We have considered
that general relativity might be an emergent feature of an underlying
quantum theory of a similar nature to that describing condensed matter
system as liquid Helium. In this case (some other cases can also be
imagined), it appears reasonable to consider the temperature of the
Universe as something independent (at least in a first approximation)
of the specific characteristics of the emergent 4-dimensional
geometry. This characteristic put us outside the realm of 
standard general relativity.

Taking this perspective, we have calculated the free energy for 
static Universes of the Einstein type filled with a gas of photons
at a fixed temperature. We have seen that the free energy has 
an extremum at the Einstein point and that this extremum correspond to 
a local minimum. Therefore, thermodynamically speaking the Einstein
state for the Universe would be stable under the stated condition.

It is not difficult to understand why this is the case. The Eddington
instability of the Einstein state is based on the following fact.
The contractive tendency of matter operates strongerly in short
scales. Instead, the expansive tendency of the cosmological constant
operates strongerly in large scales. In the Einstein point these two
tendencies are exactly balanced. However, if the universe is suddenly
made larger, the cosmological constant effect takes over and expand
further the Universe. Reciprocally, if the Universe is made smaller 
the matter dominates and makes the Universe to further contract.
However, in the case analyzed here, a sudden expansion of the 
Universe will be accompanied by the introduction of more photons
in the system in order to keep the temperature constant in the now larger
volume. This increase on the amount of matter completely 
counterbalance the cosmological constant tendency making the 
Universe to contract back to its initial state. 

 We also know that the Einstein point is unstable in standard
general relativity to homogeneous but anisotropic perturbations of the
Bianchi type IX. Taking our point of view, we have calculated the
behaviour of the Einstein point inside the general class of static
Bianchi model of type IX. We have also found that the Einstein point
is stable in our approach.
 
In summary, in the standard general relativistic view point, the
radiation dominated Einstein state is unstable under global
perturbation of the scale factor and also from homogeneous but anisotropic
perturbations. However, from the point of view of
emergent gravity these instabilities are not present.

\section*{Acknowledgements}

 We thank A.A. Starobinsky for very useful comments.  CB is supported by
the Education and Science Council of  Junta de Andaluc\'{\i}a, Spain. 
GEV  is supported by the Russian Foundations for Fundamental Research, 
by the Russian Ministry of Education and Science through the Leading
Scientific School grant $\#$2338.2003.2 and through the Research Programme
"Cosmion". This work is also supported by ESF COSLAB Programme.


\end{document}